\documentclass{PoS}

\def\lsim{\,\lower2truept\hbox{${< \atop\hbox{\raise4truept\hbox{$\sim$}}}$}\,}
\def\gsim{\,\lower2truept\hbox{${> \atop\hbox{\raise4truept\hbox{$\sim$}}}$}\,}

\title{Sunyaev-Zeldovich and Cosmic Microwave Background }

\ShortTitle{Sunyaev-Zeldovich and Cosmic Microwave Background }

\author{\speaker{Carlo Burigana}%
         \thanks{{\it Dipartimento di Fisica, Universit\`a degli Studi di Ferrara, 
via Saragat 1, I-44100 Ferrara, Italy}}
\\
        INAF-IASF Bologna, via Gobetti 101, I-40129 Bologna, Italy\\
        E-mail: \email{burigana@iasfbo.inaf.it}}

\abstract{
Since its original formulation the Sunyaev-Zeldovich (SZ) effect has been
recognized as a ``powerful laboratory'' for our comprehension 
of physical processes in cosmic structures and to derive crucial
information on some general properties of the universe.
After a discussion of the fundamental concepts and of some 
well established applications of the SZ effect 
towards galaxy clusters, I will focus on dedicated themes related 
to the SZ effect and other features in the cosmic microwave
background (CMB) of particular interest in the view of the extremely high
angular resolution observations achievable in the future with the 
Square Kilometre Array (SKA). 
SKA will allow the mapping of the thermal and density 
structure of clusters of galaxies at radio and centimetre bands
with unprecedented resolution and sensitivity 
and with an extremely accurate control of 
extragalactic radio source contamination.
The signatures from SZ effects and free-free emission
at galactic scales and in the intergalactic medium probe
the structure evolution at various cosmic times. 
The detection of these sources and their
imaging at the high resolution and sensitivity achievable 
with SKA will greatly contribute 
to the comprehension of crucial cosmological and astrophysical aspects, 
as the physical conditions of early ionized halos, quasars
and proto-galactic gas.
The spectacular improvement in our understanding of the properties 
of extragalactic radio sources at very faint fluxes 
achievable with SKA will allow to accurately model their contribution 
to the diffuse radio background, 
greatly contributing to the interpretation of 
next generation of CMB spectrum experiments devoted to probe the 
thermal plasma history at early times.
}

\FullConference{First MCCT-SKADS Training School\\
                 September 23-29, 2007\\
                 Medicina, Bologna, Italy}

\begin{document}

\section{Introduction}
\label{intro}

The thermal plasma in the intergalactic and intracluster medium and at
galactic scales leaves imprints on the cosmic microwave background (CMB) 
through the Thomson scattering
of CMB photons on  hot electrons (Sunyaev-Zeldovich effect) 
and the free-free emission.
Since its original formulation the Sunyaev-Zeldovich (SZ) effect 
has been recognized as a ``powerful laboratory'' for our comprehension 
of physical processes in cosmic structures and to derive crucial
information on some general properties of the universe. 
The characterization of the free-free emission distortion in many astrophysical
and cosmological contexts may suffer of an intrinsic uncertainty because of the 
relatively strong dependence of this effect on matter density, typically implying  
a certain model dependence in the estimates of its amplitude. In spite of this,
the theoretical predictions of several models could be probed through
new measures at radio and centimetre wavelengths.

After a discussion of the fundamental concepts 
(Sect.~\ref{kompaneets}) 
and of some 
well established applications 
(Sect. \ref{sz_clust}) 
of the SZ effect 
towards galaxy clusters, I focus on dedicated themes related 
to the SZ effect and other features in the CMB
of particular interest in the view of the extremely high
angular resolution observations achievable in the future with the 
Square Kilometre Array (SKA) 
(Sect.~\ref{SZ_cluster_SKA} and Sect.~\ref{small_scales_SKA}).
In fact, although not specifically devoted to CMB studies, because of its 
high resolution and the limited high frequency coverage, right
the extreme sensitivity and resolution of SKA may be fruitfully 
used for a detailed mapping of these effects on dedicated sky areas.
The SKA highest frequencies ($\simeq 20-30$~GHz), that will be implemented 
during the third phase of the project \cite{ska_project}, are, of course, the 
most advantageous in particular for the study of SZ effects 
because of the steeper decrease of the synchrotron radio emission 
with the frequency.

For sake of conciseness, I avoid in this work to include figures 
as well as to repeat a complete reporting of the derivations 
of some fundamental equations, focussing, on the contrary,
to their physical meaning.
They are in fact available in the presentation uploaded  
at the School web 
site\footnote{http://www.ira.inaf.it/$\sim$school\_loc/presentations/PDF/Carlo\_Burigana.pdf} 
(you can also ask the presentation to the author).
I will often refer in the following to the appropriate presentation slide number.

\section{Kompaneets equation and imprints on the CMB}
\label{kompaneets}

Under general condition, the evolution of CMB photon occupation number, $\eta(\nu)$ (at the 
frequency $\nu$), which determines the CMB photon energy density 
$\epsilon_{\rm CMB} = 8\pi h_P / c^3 \int \eta(\nu) \nu^3 d\nu$, 
is described by the kinetic (Boltzmann) equation; here $h_P$ is the Planck constant
and $c$ the speed of light. Various 
physical processes 
can contribute to the evolution of $\eta$. In many cosmological and astrophysical contexts
the Compton scattering is the most efficient interaction in the plasma able to exchange 
energy between (baryonic) matter and radiation. From fundamental electromagnetism concepts
one can derive the Compton cross-section and the energy exchange between electron and photon
in a single scattering (see e.g. \cite{Rybicki_Lightman_1979}; see slides 4--6). In the 
Thomson limit ($\epsilon/m_ec^2 \ll 1$, where $\epsilon$ is the photon energy before 
scattering and $m_e$ the electron mass; $m_ec^2 \simeq 511$~keV) 
the photon energy $\epsilon_1$ after scattering is not so different from the original one and 
the differential cross-section is almost isotropic. From the energy-momentum conservation law
in the electron rest frame it is possible to derive the photon dimensionless energy, 
$x_1=h_P\nu_1 / k_BT_e$, after scattering as function of the photon dimensionless energy, $x$,
before scattering, of the electron energy and momentum, and of the photon directions before 
and after scattering; here $T_e$ is the electron temperature and $k_B$ the Boltzmann constant. 
In general, in the electron rest frame, $x_1 < x$, or, in other words,
the photon gives a fraction of its original energy to the electron. The conclusion is 
different passing to the laboratory frame: in this case, because of Doppler effect, 
the photon energy can be increased of decreased after the scattering, depending on the 
relevant electron and photon motion directions. 
The number of scatterings per unit time depends on the electron density, $n_e$, and on the 
photon ($c\vec{n}$) and 
electron ($\vec{v}$) velocities

\begin{equation}
{dN \over dt} = \sigma_T n_e c [1-(\vec{v}/c)\cdot \vec{n}] \, ;
\end{equation}

\noindent
here $\sigma_T$ is the Thomson cross-section and $t$ is the time.

Averaging over an ensamble of electrons at a temperature $T_e$
and using the semi-isotropy of scattering holding in the Thomson limit,
the average change of the photon energy per unit time due to the Doppler 
effect can be expressed by (see slide 6)

\begin{equation}
{<d h \nu >  \over dt}_D = \sigma_T n_e c h_P \nu {k_BT_e \over m_ec^2} \, .
\end{equation}

The kinetic equation describes as $\eta(\nu)$ changes because of the 
contribution from photons at different frequencies and 
the migrations of photons from the frequency $\nu$ to frequencies different 
from $\nu$ because of Compton scattering (see slide 8).
It is general but at the same time very difficult to solve. For this reason,
various approximations of it have been proposed in the literature according to the
assumptions applicable to the considered astrophysical or cosmological context.
The most famous approximation has been derived by Kompaneets many years before its
publication \cite{kompaneets}. It is a second order approximation of
the kinetic equation in the dimensionless energy difference $\Delta = x_1 - x$
in the Thomson limit for the differential cross-section 
and for a Maxwellian distribution of electrons (see slide 8). These two assumptions greatly 
simplify 
the second order serie expansion allowing to identify two contributions to the time evolution
of the photon occupation number, 
$\frac{\partial\eta}{\partial t}$: a secular shift term, $\propto \Delta$, and a random walk 
 term, $\propto \Delta^2$. Thanks to the semi-isotropy of the Thomson differential 
cross-section
the latter term can be easily computed directly (see slide 9). The derivation of the 
secular shift 
term can instead be performed using the fundamental property of scattering, i.e. the photon 
number conservation. It sets a constraint on the photon flux on a spherical surface of the 
phase space at a given frequency, $j(x)$, equivalent to the relation 

\begin{equation}
\frac{1}{x^2}\frac{\partial j(x)}{\partial x} = \frac{\partial\eta}{\partial t} \, .
\label{etadij}
\end{equation} 

\noindent
Since no products of the first and second derivative of $\eta$ with 
respect to $x$ appear in $\frac{\partial\eta}{\partial t}$, then
$\frac{\partial\eta}{\partial x}$
can appear only linearly in the function $j$.
In addition, 
at thermal equilibrium, when the CMB spectrum 
is a blackbody (BB), $j=0=\frac{\partial\eta}{\partial t}$. This sets the form of $j(x)$ to be a 
product of a function of $\eta$ by an unknown function $g(x)$. 
$\frac{\partial\eta}{\partial t}$ can be formally expressed by 
Eq.~(\ref{etadij}) and by the above second order serie expansion. Therefore, 
when the random walk term is explicited, their equality and, in particular, the 
equality between the coefficients in front of the first and second derivative of $\eta$ with
respect to $x$ allows to determine the form of the function g(x) and then the 
expression of the secular shift term (see slide 10). One then arrives to the
Kompaneets equation (for the scattering Compton term alone).
In cosmological context, it is important to note that the dimensionless photon energy 
$x=h_P\nu / k_BT_e$ is in general not redshift invariant, depending on the 
redshift dependence of $T_e$. It is then more advantageous to introduce
a different dimensionless photon energy defined by
$x=h_P\nu/[k_BT_0(1+z)]$ where, independently of the exact form of the photon occupation
number at the present time $t_0$, 
$a_{BB}T_0^4$ gives the present CMB energy density, 
$\epsilon_{CMB,0}$, $a_{BB}$ being the
BB constant. It is useful to introduce a ``nominal'' 
CMB energy density $\epsilon_{\rm CMB} = a_{BB} T_{\rm CMB}^4 \simeq 4.2 \times
10^{-13}(1+z)^4\,\hbox{erg}\,\hbox{cm}^{-3}$, where
$T_{\rm CMB}= T_0(1+z)$ is a ``nominal'' CMB temperature\footnote{``Nominal'' 
refers here to the case in which the CMB energy density
at any epoch scales simply as $(1+z)^4$, i.e. in the absence of energy exchanges on the 
cosmic plasma.}. 
In this way, $x$ is redshift invariant (since also $\nu$ scales as $1+z$ during 
cosmic expansion).
Adopting this last choice of $x$, the Kompaneets equation can be written as
\begin{eqnarray}\label{eq:kompaneets_C}
\frac{\partial\eta}{\partial t} = \left({\frac{\partial\eta}{\partial t}}\right)_C =
\frac{1}{\phi}\frac{1}{t_C}
\frac{1}{x^2}\frac{\partial}{\partial x}\left[x^4\left[\phi
\frac{\partial\eta}{\partial x}+\eta(1+\eta)\right]\right] \, ;
%\\
%\nonumber\\
%&+&\left[K_{BR}\frac{g_{BR}}{x_e^3}e^{-x_e}+K_{DC}\frac{g_{DC}}{x_e^3}+K_{CE}
%\delta(x_e-x_{e,CE})\right]\left[1-\eta(e^{x_e}-1)\right] \, , \nonumber
\end{eqnarray}

\noindent
here $t_C=m_ec^2/[k_BT_e(n_e \sigma_T c)]$ 
is the gas cooling time by Thomson scattering
and $\phi=T_e/[T_0(1+z)]$
is a dimensionless electron temperature. 
Three terms can be identified in Eq.~(\ref{eq:kompaneets_C}): 
ordinary and induced Compton scattering, respectively 
proportional to $\eta$ and $\eta^2$, that tend to move photons towards low frequencies
and inverse Compton scattering, the only one involving the second derivative 
of $\eta$, that tends to move photons towards high frequencies \cite{DD77}. 
It is important to remember that Eq.~(\ref{eq:kompaneets_C}) needs to be coupled
with the appropriate (differential equation describing 
the) time evolution of the electron temperature (see slides 7 and 24) 
in order to avoid to introduce significant errors \cite{BDD91a} in its 
solution\footnote{The dependence of the Compton equilibrium electron temperature 
\cite{peyraud,ZL68} on integrals of $\eta$ increases the computational  
complexity of the (numerical) solution of Eq.~(\ref{eq:kompaneets_C}) \cite{BDD91a}.}.

At this point it is important to remember the limits of validity of this equation
(see slide 12): a large number of scatterings is assumed (otherwise it has no meaning
to compute averages); non relativistic electrons and relatively soft photons are assumed;
the photon occupation number shape should be smooth (otherwise 
$\frac{\partial\eta}{\partial x}$ could be much larger than $\eta$, and so on
for higher order derivatives, and 
a serie expansion is no longer adequate). Finally, at least in this form,
the Kompaneets equation assumes isotropic distributions of electrons and photons.

Note that, in general, for a Bose-Einstein 
(and, obviously, also for the particular case of a Planckian)
distribution $\frac{\partial\eta}{\partial t} = 0$ 
(see slide 15 for the fundamental properties of these distributions): 
in other words, if there is enough time
to allow a sufficient number of scatterings,
under the action of Compton scattering alone the photon occupation number tends 
towards a Bose-Einstein (BE) distribution which then represents the (general) kinetic 
equilibrium solution.
Typically, for CMB photons, this condition is satisfied at high redshifts at least at 
high frequencies, where other radiative processes are not so important.
The chemical potential\footnote{In this context, the dimensionless chemical potential
is typically used.} $\mu$ of the BE distribution is related to the fractional
energy exchange occurred between matter and radiation associated to the 
dissipation process at the origin of the deviation from the full equilibrium 
BB spectrum \cite{SZ70,DD77}:
\begin{equation}
\mu\simeq 1.4 \Delta\epsilon/\epsilon_i \, ;
\end{equation}

\noindent
the case of small distortions 
(or small energy exchanges 
$\Delta\epsilon/\epsilon_i \ll 1$, at least in the case of dissipation 
processes with negligible photon production) 
is considered here for simplicity; $\epsilon_i$ is the CMB energy density 
before energy exchange. 
This result can be easily derived comparing the photon energy density and number density
of the original BB spectrum and of the distorted BE spectrum.
It is important to note that, observationally, 
COBE/FIRAS data set stringent limits to the amplitude of spectral distortions
at redshifts lower than $\approx 10^6$, while these limits are significantly relaxed 
at earlier epochs \cite{fixsen96,SB02} (see slides 28--29 and 31). At $z \gsim z_{therm}$, $z_{therm}$
being the thermalization redshift (see slide 30) \cite{BDD91a}, the limits on 
$\Delta\epsilon/\epsilon_i$ are set by the cosmological nucleosynthesis
theory, that, in order to explain chemical abundances, implies an early comoving 
photon energy density not too far from the present one.

A very different solution can be found under two different assumptions.
If $\phi \gg 1$ (hot electrons) the inverse Compton term 
dominates over ordinary and induced Compton terms. Neglecting these two terms, the
Kompaneets equation becomes equivalent to the heat diffusion equation 
(see e.g. \cite{tricomi}) 
and admits an exact analytical solution 
described by a superposition of blackbodies \cite{ZeldovichSunyaev1969,ZIS72}. 
Note that (see slide 16) 
for small deviations from an initial BB spectrum the Kompaneets equation
is equivalent to the inverse Compton dominated case except for a further
factor $(1-\phi_i/\phi)$ where $\phi_i$ is the initial electron and (BB) radiation
dimensionless temperature (corresponding to a physical temperature $T_i$). 
A simple solution, the so-called Comptonization solution, can be found in this case:
\begin{eqnarray}\label{eq:solbassiz_H}
\eta(x,\tau) \simeq
%\eta^H(x,\tau)
%= 
\eta_i+u\frac{x/\phi_i\exp(x/\phi_i)}
{[\exp(x/\phi_i)-1]^2}\left(\frac{x/\phi_i}{\tanh(x/2\phi_i)}-4\right) \,
,
\end{eqnarray}

\noindent
where $\eta_i$ is the initial distribution function and
$u$ is the (evolving) Comptonization parameter (see e.g. \cite{BDD95})
\begin{eqnarray}\label{eq:upar1}
u(t)=u(z) = \int_{t_i}^t\frac{\phi-\phi_i}{\phi}\frac{dt}{t_C}=
\int_{1+z}^{1+z_i}(\phi-\phi_i)\left(\frac{k_BT_r}{m_ec^2}\right)n_e\sigma_T
t_{exp}\frac{d(1+z^{\prime})}{1+z^{\prime}} \, ;
\end{eqnarray}

\noindent
here $t_{exp}=a/(da/dt)$ is the cosmic expansion time.
Obviously, the case of hot electrons is directly obtained from this equation
in the limit $\phi \gg \phi_i$. This formula is an excellent approximation
of the exact solution.
By integrating $u$ over the whole relevant energy dissipation phase
one gets the ``usual'' Comptonization parameter
related to the whole fractional energy exchange by the
well known expression \cite{ZeldovichSunyaev1969} 
\begin{equation}
u\simeq (1/4)\Delta\epsilon/\epsilon_i \, ;
\end{equation}

\noindent
again, the case of small distortions 
%(or small energy exchanges 
%$\Delta\varepsilon/\varepsilon_i \ll 1$, at least in the case of dissipation 
%processes with negligible photon production) 
is considered for simplicity in the above equation.
This result can be easily derived considering how the photon energy density
changes because of the evolution of $\eta$ (see slide 13).

\noindent
Note that in the Rayleigh-Jeans (RJ) region Eq.~(\ref{eq:solbassiz_H}) predicts a lowering
of the photon occupation number (see slide 16) and of the brigthness temperature
(see slides 17). This effect decreases at increasing frequency up to 217~GHz
where it vanishes. Then, for a further frequency increasing the formula predicts 
an increasing signal fractional excess with respect to the original BB.
In other words, inverse Compton scattering tends to create a photon distribution
with more (respec. less) high (respec. low) frequency photons than those of the original BB.
In cosmological context this kind of departure from a BB (i.e. of CMB spectral 
distortion) is expected at relatively late epochs (at early ones there is time enough to 
achieve the kinetic equilibrium) in the 
presence of a diffuse medium at temperature larger
than that of the CMB ($T_{matter} = T_e > T_0 (1+z)$; see slides 32--33), but also under
many physical conditions (see e.g. \cite{DD77}). 
Note that, at least in principle, a cooling process ($T_{matter} = T_e < T_0 (1+z)$)
could produce a negative Comptonization distortion \cite{stebb_silk} (see slides 32 and 36).
A remarkable 
example\footnote{
In the past, before of its explanation 
in terms of integrated contribution by discrete sources,
the cosmic X-ray background (see slide 34) has been interpreted in terms of 
bremsstrahlung radiation by hot diffuse medium. In this framework,
a substantial CMB Comptonization distortion was predicted, but right 
the stringent upper limit to this kind of CMB distortion posed by COBE/FIRAS 
sets the most stringent upper limit to the fraction of cosmic X-ray background 
that could be produced by diffuse hot gas
(see slide 35).} 
of (positive) Comptonization distortion is that 
associated to the cosmological reionization
that affects the CMB 
both in anisotropies at large and small scales and in 
the spectrum
(see slide 81; see \cite{ciardischool}; see also Sect.~\ref{cmbspec}).
For instance, the reionization redshift (see slide 82)
identified by 
the NASA {\it Wilkinson Microwave Anisotropy 
Probe}\footnote{http://lambda.gsfc.nasa.gov/product/map/current/}
(WMAP) satellite 3-yr data 
%\cite{kogutetal03}
is consistent with minimal reionization models \cite{schneider}
predicting a Comptonization distortion  
with $\Delta \epsilon / \epsilon_i \simeq 4u \approx 4-7 \times 
10^{-7}$ \cite{BPSSCF07}. 

At low frequencies, other radiative processes can not be neglected and the corresponding
terms have to be included in the (complete) Kompaneets equation:

\begin{equation}\label{eq:Kompaneetsformal}
\frac{\partial\eta}{\partial t}=\left(\frac{\partial\eta}
{\partial t}\right)_{\Lambda}+\left(\frac{\partial\eta}
{\partial t}\right)_{\Gamma}=\sum\Lambda_i+\sum\Gamma_i \, ;
\end{equation}
here $\sum\Lambda_i$ takes into account processes that does not change the
photon number (typically, the Compton scattering)
and $\sum\Gamma_i$ takes into account
photon production/absorption processes (see slides 20--22).\\%
Three mechanisms certainly operate in the cosmic plasma: bremsstrahlung (BR) or free-free (FF)
\cite{KL61,Rybicki_Lightman_1979}, 
double (or radiative) Compton (DC) \cite{gould}, and cyclotron (CE) emission.
Recently, \cite{zizzo_burigana} demonstrated that, for realistic values of 
cosmic magnetic field, the cyclotron process never plays an important
role for (global) CMB spectral distortions when ordinary and stimulated emission
and absorption are properly taken into account and CMB realistic distorted spectra,
as discussed in the following, are considered. In fact, the cyclotron term may be 
significant, in the case of deviations of $\eta$ from the BB distribution at the electron
temperature, only at very long wavelengths, corresponding to the cyclotron frequency, 
where, during the formation of a spectral distortion, FF and DC are able to keep
$\eta$ extremely close to the BB equilibrium. These processes are in fact very efficient
at long wavelengths because of their $\approx x^{-3}$ dependence. So, at high redshifts the BE
solution needs to be modified introducing a frequency dependence in the chemical 
potential, $\mu=\mu(x)$, which vanishes at very long wavelengths \cite{SZ70,DD80}. 
The resulting spectrum
(see slide 25) has then a minimum in equivalent thermodynamic 
temperature\footnote{$T_{th}$ is defined as the temperature of a BB with the same 
$\eta(\nu)$ as that under consideration, i.e. such that 
$\eta(\nu) = 1/[\exp(h_P\nu/(k_B T_{th}(\nu)))-1]$. 
Only for $\eta=\eta_{BB}$, $T_{th}$ is independent of $\nu$, while for an arbitrary photon 
occupation number $T_{th}$ is frequency dependent, $T_{th}=T_{th}(\nu)$. 
In the following $T_{th}$ will be 
called also brightness temperature and denoted by $T_{br}$. Note that the brightness temperature,
$T_{br}$, is often used in the literature also with the meaning of antenna temperature 
$T_{ant}$, defined as the temperature of a BB with the same
$\eta(\nu)$ as that under consideration but in the RJ limit, 
i.e. $\eta(\nu) = 1/[h_P\nu/(k_B T_{ant}(\nu)]$. 
Obviously, $T_{ant} \simeq T_{th}$ in the RJ limit.},
$T_{th}$,
at a wavelength depending on the baryon density \cite{DD80,BDD91a}. Note that DC (respec. FF) is 
more important that FF (respec. DC) at high (respec. low) redshifts ($z$ higher 
(respec. lower) than $\approx 10^5$). In general, if there is enough time, the long 
wavelength photon production by DC and FF combined to the shifting of produced photon 
to higher frequencies thanks to  Compton scattering tends to reduce the amplitude
of a possible early spectral distortion, being at limit able to fully 
re-establish a BB (see slides 28--31), i.e. to thermalize the CMB spectrum \cite{BDD91a}. 

The BE and Comptonization like distortions discussed above (properly modified
with respect to the pure BE and Comptonization distributions because of the
effect of DC and FF) describe distorted spectra at early and late
epochs, respectively (see slide 18). For processing possibly occurred at 
intermediate epochs 
a distorted spectral shape intermediate between these two
cases is predicted (see slide 26). In particular, for processes at relatively late 
epochs a plateau 
(in $T_{th}$) is produced in the RJ region, with a 
brigtness temperature decrement related also to the 
process epoch\footnote{It can be formally derived with a further approximation 
of the Kompaneets equation in terms of small deviations
from the Comptonization spectrum.} \cite{BDD95} (see slide 26, panel b).

At relatively low redshifts, FF, instead of thermalizing the spectrum to a BB at the 
electron temperature (this is achieved in any case at extremely low frequencies), 
can produce the so-called free-free (low frequency) distortion \cite{BDD95}. 
It is easy to verify (see slide 23) that, in the RJ region, 
the $x^{-3}$ dependence of FF and the weak frequency dependence of the FF Gaunt 
factor implies a fractional increase of the equivalent thermodynamic temperature
almost proportional to $x^{-2}$, characterized in amplitude by a simple parameter,
the so-called FF distortion parameter, $y_B$. A remarkable excess
can be produced at centimetre and decimetre wavelengths (see slide 33) -- or, in principle,  
a decrement in the case of cooling processes (i.e. negative distortions; see slide 36). 

\section{Thermal and kinetic SZ effect towards galaxy clusters}
\label{sz_clust}

The Comptonization solution described in the previous section is at the basis 
of the well known (thermal) SZ effect. The form of 
Eqs.~(\ref{eq:solbassiz_H}) and (\ref{eq:upar1}) are appropriate to the case 
of CMB global distortions, but it easy to rewrite these equations in the case of the 
interaction of CMB photon with hot gas in a cluster of galaxies (or in a primeval halo 
or, at smaller scales, in the hot medium surrounding a galaxy). It is enough to replace 
the integral 
in $n_e c dt$ over the relevant cosmic time with an integral in
$n_e dl$ over the cluster along the line of sight in Eq.~(\ref{eq:upar1}) (see slide 37).

The SZ effect due to the scattering of CMB photons 
with hot electrons in galaxies 
and clusters of galaxies is typically expressed 
in terms of a frequency dependent change
in the CMB brightness (see slides 39--40). If the hot electron gas is globally  
at rest with respect to the observer,  only the thermal 
SZ effect \cite{sz72} (see also \cite{rephaeli95}) 
will be present; 
differently, a bulk 
peculiar motion, $V_r$, of the hot electron gas produces  a kinetic
SZ effect (see slide 38). In the RJ region the first effect 
produces a decrement of the surface brightness,
$\Delta I_{th}$, towards the cluster. The second effect
produces either a decrement or an increment, $\Delta I_k$,
depending on the direction of the cluster velocity with respect to 
the observer.
Neglecting relativistic corrections: 

\begin{equation}
\Delta I_{th} = I_0 y g(x)
\end{equation}
and   
\begin{equation}
\Delta I_k = -I_0 (V_r/c) \tau_e h(x) \, ,
\end{equation}
where   
$I_0=(2h_P/c^2)(k_B T_{CMB}/h_P)^3$. Here 
\begin{equation}
\tau_e=\int n_e \sigma_T dl
\label{tau_electr}
\end{equation}
and 
\begin{equation}
y=\int (k_BT_e/m_ec^2) n_e \sigma_T dl
\end{equation}
are respectively the Thomson optical depth and the Comptonization 
parameter\footnote{I prefer to use here the symbol $u$ for the ``global'' 
Comptonization parameter of a CMB distorted spectrum and the symbol $y$ for the 
``local'' Comptonization parameter characterizing the SZ effect.} 
\cite{ZeldovichSunyaev1969} 
integrated over the cluster along the line of sight and
\begin{equation}
h(x)=x^4e^x/(e^x-1)^2
\end{equation}
\begin{equation}
g(x)=h(x) [x(e^x+1)/(e^x-1)-4] \, .
\end{equation}
The two effects have a different frequency dependence 
that in principle allows their separation through 
multi-frequency observations (see slides 41--42). 
SKA observations in the RJ regime (where $h(x) \sim g(x) 
\rightarrow x^2$) 
can be combined with millimetric observations. 
In particular, $g(x) \simeq 0$ and $h(x)$ is maximum 
at $\sim 217$~GHz, a frequency that, exactly for this reason, is 
typically included in SZ multifrequency 
observations at millimetre wavelengths. Accurate measures of the kinetic SZ effect
can provide a crucial cosmological information on cluster velocity field (see e.g. 
\cite{benson} for results achieved with SuZIE II).

\subsection{Treatments of the thermal SZ effect beyond the Kompaneets approximation}

The above treatment of the SZ effect is based on the Kompaneets equation
and obvioulsy assumes the same hypotheses (or work conditions).

Two main lines have been followed to overcome the limits of validity
of this formalism. From one side one can search for solutions more appropriate
to the case of a small number of scatterings between CMB photons and electrons,
possibly including (at least to a given order) the Klein-Nishina correction
to the classical Thomson scattering cross-section \cite{wright,fabbri}.
From the other side, it is interesting to search for approximations of the kinetic 
equation not only up to the second order but up to a higher order,
necessary to treat the relativistic case \cite{itoh}.
These approaches are physically motivated on the basis of the high electron
temperatures often achieved in clusters of 
galaxies\footnote{Or for a diffuse very hot gas -- see also footnote 3.}
and/or by small values of the optical depth (see Eq.~(\ref{tau_electr})).

The general properties of the solutions found using these treatments
are not so different from those predicted by the original SZ treatment.
See slide 35 for an application of the method described in \cite{fabbri}: note the slightly
steeper behaviour at high frequencies found in this case. See slide 44 for
a comparison between non-relativistic and relativistic approximations \cite{RSS05}: note 
how the difference, clearly appreciable at high frequencies, is not 
so critical in the RJ region (relevant for SKA). Note also that only a 
small shift of the frequency of vanishing effect with respect to the 
non relativistic case (217 GHz) is predicted.

\subsection{Some cosmological applications of SZ effect towards galaxy clusters}

The exploitation of the cosmological applications and implications of the SZ effect
towards galaxy clusters has received a large attention from both the theoretical
and the observational point of view. The reader could refer for example to 
\cite{varenna} (and references therein) 
for exhaustive discussions (see slide 57). 
I report here on three interesting examples.

\subsubsection{Hubble constant determination}

As well known, one of the most remarkable difficulties in the accurate measure
of the Hubble constant, $H_0=[(da/dt)/a]_0$ (see e.g. \cite{weinberg}), 
which determines the universe expansion rate
at the present time, has been represented for a long time by the necessity 
to precisely calibrate various kinds of ``standard candles''. So, some 
attempts have been dedicated to find methods able to measure $H_0$ (as 
well as other cosmological parameters) circumventing this (or other) difficulty(ies). 
An interesting solution has been proposed in the '70 by \cite{CDD78,gunn78,SW78,CDD79} (see also 
\cite{partridge}). It 
involves the combination of X-ray and radio observations towards cluster of galaxies 
(see slides 46--48). 
The different dependence of bremsstrahlung, responsible 
of the cluster X-ray emission, and of SZ effect (namely the decrement 
of the brigthness temperature in the RJ region) towards the cluster
on the relevant physical and geometrical parameters allows to derive a 
simple expression for the cluster luminosity distance in terms of 
ratio between the square of the brigthness 
temperature decrement in the RJ region and the X-ray brightness, of cluster electron 
temperature, cluster (angular) core radius, and cluster redshift. Note that these 
quantities are, in principle, measurable. Therefore, a comparison with the usual
Hubble expression for the luminosity distance (see slide 48) allows to directly 
derive $H_0$. A certain degradation of the reliability of this method comes from 
the complexity of the geometrical and physical properties of clusters (see e.g.
slide 49). For example, different baryonic physics results into 
noticeable differences in the surface 
distributions and in the profiles of the X-ray brightness, projected temperature, and 
Comptonization parameter, particularly for the inner cluster regions \cite{norman} (see slide 50). 
Another remarkable difficulty of the method and, in general, of cluster modelling,
is represented by the presence of
radio sources in the cluster that, with their surface density distribution different 
from that of control fields, can affect the signal due to the ``pure'' SZ effect,
a problem recently addressed by \cite{massardi_dezotti} (see slides 51--52).

Only recently, accumulating an increasing number of clusters, this method 
to measure $H_0$ starts to provide reliable results \cite{calstrom}. 
In spite of the above difficulties, 
the large
increasing of number of clusters that will be possible to consider in the next future
(see Sect.~\ref{SZ_cluster_SKA}) and the continuous improvement in the
observation and physical modelling of clusters, will make this method very interesting
at least for an almost independent cross-check of the results achievable
through other measurements, first of all from the CMB anisotropy.

\subsubsection{Determination of cosmological parameters through statistical approaches}

The Hubble constant is only one of the parameters that determine the 
``background''\footnote{``Background'' is used here to denote the global properties
of a (``virtual'') perfectly isotropic and homogeneous universe with all 
(``virtual'') galaxies following the pure Hubble expansion flow.} 
evolution of the universe, while other parameters 
characterize the dishomogeneity  properties of the universe related to the formation
and evolution of cosmic structures. On the other hand, the
``background'' evolution of the universe plays a crucial role also for the 
structure formation process. Therefore, the study of the correlation properties
of such dishomogeneities provides information on the cosmological parameters
originally introduced to characterize the ``background'' properties of the universe.

Likely, the most useful (and used) statistical estimator of such dishomogeneities
projected on the celestial sphere is their angular power spectrum, $C_\ell$,
as function of the multipole\footnote{The multipole $\ell$ is almost 
inversely proportional to the angular scale, $\ell \simeq {180}/{\theta(^{\circ})}$.}, 
$\ell$, that, observationally, is a simple
transform of the angular (auto)correlation function \cite{peebles}. For Gaussian
distributions, they completely describe the statistical properties of the
fluctuation field. From the astrophysical side, $C_\ell$ is related to the 
redshift dependent mass function and profile of clusters. Different models for the 
mass function and for characterizing the physical profile of clusters 
imply different shape and amplitude of $C_\ell$ (see slide 53). 
In particular, it takes memory 
of some cosmological parameters, such as the density contrast
(i.e. the linear theory amplitude of matter fluctuations), $\sigma_8$, at a 
characteristic scale (typically, 8~$h^{-1}$~Mpc)  and the global matter density (see 
\cite{RSS05,arnaud} and reference therein); here
$h$ is the Hubble constant in units of
$100~\hbox{km}\,\hbox{s}^{-1}\,\hbox{Mpc}^{-1}$.
So, the comparison between the observed and the theoretically predicted
cluster distribution allows to determine such parameters or can be used with other 
cosmological data to remove the degeneracy in the determination of some
cosmological parameters, existing using a single kind of observables (see e.g. 
\cite{deg_bond1,deg_bond2,EB99}),
such as the cosmological constant (or dark energy) and the
matter (baryonic plus dark matter) density parameters (see slides 54--55).

\subsubsection{The SZ effect as a foreground for CMB studies}

Recently, a spectacular improvement in the assessment of cosmological models and in 
the determination of the cosmological parameters has been achieved thanks to the 
precise mapping of CMB anisotropies through balloon-borne, ground-based, and, in 
particular, space experiments (see e.g. \cite{BMM02}, and references therein, for a review 
on pre-WMAP observational results). 
WMAP measured CMB anisotropies 
on the whole sky in both total intensity 
(or temperature, see 1-yr results \cite{bennett03}) 
and polarization 
(see the 3-yr results \cite{jarosik07,page07}) Stokes parameters. 
A substantial 
improvement is expected in the next years from the forthcoming 
ESA {\it Planck}\footnote{http://www.rssd.esa.int/planck} 
satellite \cite{mandolesi,puget,planckcoll}. 

The SZ effect from the ensamble of the clusters at various redshifts and sky 
positions constitutes a source of foreground for these projects, particularly 
remarkable at small scales (or large $\ell$). The SZ effects towards the brightest 
clusters can be directly identified on the CMB anisotropy maps and can be subtracted 
almost accurately from the fluctuation map during data analysis. 
On the contrary, those below a certain threshold can not be extracted from the fluctation 
field and contribute to the overall angular power spectrum (see slide 56) \cite{hansen,dolag}.
Therefore, the contribution of SZ effect from clusters to the overall $C_\ell$
needs to be accurately modelled and subtracted to extract the $C_\ell$
corresponding primary CMB fluctuations. These tasks, already carried out in the WMAP
data analysis, are more relevant, and complex, in the case of the next {\it Planck} 
data analysis since it is aimed at measuring the CMB $C_\ell$ at significantly higher 
multipoles. Clearly, the {\it Planck} wide frequency coverage, from 30~GHz to 
857~GHz, i.e. at frequencies both below and above $\simeq 217$~GHz where the SZ 
effect vanishes, will allow to detect both the decrement and the excess 
of the CMB brightness temperature. This will greatly help the identification of this 
source of foreground, providing at the same time a rich information on cluster 
physics (see also the next section).

\subsection{SKA perspectives}
\label{SZ_cluster_SKA}

Many thousands of galaxy clusters
can be identified on the whole sky by XMM ($\sim 10^3$), 
{\it Planck} ($\sim 5000$), and SDSS ($\sim 5 \, 10^5$).
%Goto T., et al. 2002, AJ, 123, 1807  per SDSS su 350 gradi quadrati 
The typical angular sizes of galaxy clusters range from $\sim$~arcmin to
few tens of arcmin.

The SKA sensitivity and resolution mainly depends on the used array 
collecting area (see slides 60--63).
By considering a frequency band of $\simeq 4$~GHz at 20~GHz,
the whole instrument collecting area 
will allow to reach a (rms) sensitivity of $\simeq 40$~nJy 
in one hour of integration with 
an angular resolution of $\simeq 1$~mas
(considering $\simeq 3000$~km maximum baseline).
By using only about 50\% of the collecting area
within $\simeq 5$~km, the (rms) sensitivity
in one hour of integration 
is $\simeq 80$~nJy with a  resolution of $\simeq 0.6''$
\cite{ska_project,buriganaetal04a}. 

The major role on the study of the SZ effects
towards galaxy clusters will be obviously played by 
dedicated telescopes operating at $\simeq$ arcmin resolutions with 
frequency coverages up to $\simeq$ millimetric wavelengths
\cite{jones} and, at higher resolutions, by ALMA\footnote{http://www.alma.cl/}.
On the other hand, with the 50\% of the SKA collecting area it will be possible 
to accurately map the 
SZ effect  \cite{subra_ekers}
of each considered cluster, particularly at moderately high redshifts, 
with the unprecedented precise subtraction of discrete radio sources 
achievable with SKA.
% (see also Sect.~\ref{cmbspec}).
The combination with accurate X-ray images of the cluster bremsstrahlung emission
will allow to accurately map the thermal and density 
structure of the gas in galaxy clusters, thanks to the different dependences
of these two effects on matter density and temperature.
Of particular interest in this respect will be the observations
expected by the wide field imager (WFI) on board the 
X-ray Evolving Universe Spectroscopy (XEUS) 
satellite\footnote{http://sci.esa.int/science-e/www/area/index.cfm?fareaid=25},
recently 
selected by ESA as new candidate for possible future scientific 
missions\footnote{http://sci.esa.int/science-e/www/area/index.cfm?fareaid=100},
designed to reach a resolution of $0.25''$ on a FOV of $5'-10'$
(see slides 64--67).

As mentioned above, it is also possible to study 
the SZ effect (both thermal and kinetic) from clusters in a 
statistical sense, namely through its contribution to the 
angular power spectrum
of the CMB secondary anisotropies 
(see slides 68--70; see also Appendix A in \cite{buriganaetal04a}).  
This topic has been investigated in several papers 
(see, e.g., 
\cite{ostrikervishniac,vishniac,gnedin2000,springel2001,dasilva2001,ma_fry_2002}). 
At sub-arcmin scales (i.e. at multipoles $\ell \gsim 10^4$) 
secondary anisotropies from thermal 
(more important at  $\ell \lsim {\rm few} \times 10^4$)
and kinetic 
(more important at  $\ell \gsim {\rm few} \times 10^4$)
SZ effect dominate over CMB primary anisotropy 
(see slide 78)
whose
 power significantly 
decreases at multipoles $\ell \gsim 10^3$
because of photon diffusion (Silk damping effect \cite{silk68}).
Their angular power spectrum 
at $\ell \sim 10^4-10^5$ ($\approx 10^{-12} - 10^{-13}$ in terms
of dimensionless $C_\ell \ell (2 \ell+1)/4\pi$)
could be in principle investigated  
with the sensitivity achievable with SKA (see slides 71--77; see \cite{knox95}
and Appendix A in \cite{buriganaetal04a}).  
On the other hand, at the SKA resolution and sensitivity 
the contribution to fluctuations from foreground sources 
(both diffuse radio emission, SZ effects, and free-free
emitters) at galaxy scales likely dominates over the SZ effect from 
clusters (see slides 79--80).

\section{Imprints on the CMB at small scales: perspectives from the SKA}
\label{small_scales_SKA}

In the view of the extreme resolution of the SKA, it is natural to expect
that it will provide a particular improvement for our understanding of 
astrophysical processes able to produce SZ effect or other signatures in the CMB at 
very small angular scales. I will report here on three specific topics, 
from lower to higher redshifts, of particular
interest for extragalactic astrophysics and cosmology, properly 
revised in the context of the SKA scientific case \cite{buriganaetal04a}
and updated in some cases to account for WMAP 3-yr results and recent analyses.

\subsection{Thermal SZ effect at galaxy scale}
\label{sz_gal}

The proto-galactic gas
is expected to have a large thermal energy content, leading to a
detectable SZ signal, both when the protogalaxy collapses with the
gas shock-heated to the virial temperature 
\cite{ReesOstriker1977,WhiteRees1978}, and in a later phase as
the result of strong feedback from a flaring active nucleus
(see slide 83; see, e.g.,  
\cite{Ikeuchi1981,Natarajanetal1998,NatarajanSigurdssson1999,Aghanimetal2000,Plataniaetal2002,Lapietal2003}. 
The astrophysical 
implications of these scenarios have been investigated by \cite{dezottiSZ}.

A fully ionized gas  with a thermal energy density
$\epsilon_{\rm gas}$ within the virial radius
\begin{eqnarray} R_{\rm vir} 
%&
=
%& 
\left({3 M_{\rm vir}\over 4\pi
\rho_{\rm vir}}\right)^{1/3}  
%\\
%& 
\simeq & 1.6\, 10^2 h^{-2/3} (1+z_{\rm vir})^{-1} 
%\nonumber \\
%&& 
\left({M_{\rm vir}\over 10^{12} M_\odot}\right)^{1/3}\
\hbox{kpc} \nonumber \, 
\end{eqnarray}
transfers to the CMB an amount 
%\begin{equation}
$\Delta \epsilon \simeq  (\epsilon_{\rm gas} / t_C)   2 (R_{\rm vir}/c)$ 
%\ \label{Delta}
%\end{equation}
%
of thermal energy through
Thomson scattering producing a Comptonization parameter 
%characterizing the amplitude of
%the Sunyaev-Zeldovich effect due to this gas
%, can be estimated as
\cite{ZeldovichSunyaev1969} 
$y \simeq (1/4) \Delta \epsilon / \epsilon_{\rm CMB}$ (see slide 84).
%\label{y2}
%\end{equation}
%
Here 
$\rho_{\rm vir}\simeq 200 \rho_u$, 
$\rho_u=1.88 \times 10^{-29}h^2(1+z_{\rm vir})^3\,\hbox{g}\,\hbox{cm}^{-3}$ 
is the mean density of the
universe at the virialization redshift $z_{\rm vir}$.

Assuming the binding energy ($E_{\rm b, gas}= M_{\rm
gas} v_{\rm vir}^2$, $v_{\rm vir}=162 h^{1/3}(1+z)^{1/2}(M_{\rm
vir}/10^{12}M_\odot)^{1/3}$ km~s$^{-1}$ being the
circular velocity of the galaxy at its virial radius 
\cite{Navarroetal1997,Bullocketal2001})
to characterize the thermal energy content of the
gas, $E_{\rm gas}$,
the amplitude of
the SZ dip observable in the RJ 
region\footnote{Note that $\left|\Delta T\right|_{\rm RJ} \simeq 2 u T_0$ or 
$\left|\Delta T\right|_{\rm RJ} \simeq 3 u T_0$ respectively when
$\Delta T_{\rm RJ} = T_{\rm RJ} - T_{i,0}$ or $\Delta T_{\rm RJ} = T_{\rm RJ} - T_0$
is considered (see slide 17), being $T_{i,0}<T_0$ in the presence of a global heating 
process. This consideration applies to the case of CMB global Comptonization distortions. 
In this context, since we are considering a (local) SZ effect
characterized by a Comptonization parameter $y$, with fully negligible global energy exchange,
$\left|\Delta T\right|_{\rm RJ} \simeq 2 y T_0$.}
can be written as (see slide 85):
\begin{eqnarray}
\left|\Delta T\right|_{\rm RJ} 
%&
=
%& 
%2yT_{\rm CMB} 
2yT_0
%\\
% & 
\simeq 
%&  
1.7
\left({h\over 0.5}\right)^{2} \left({1+z_{\rm vir}\over 3.5}\right)^3 
%\nonumber \\
%&& 
{M_{\rm gas}/M_{\rm vir}\over 0.1} {M_{\rm vir}\over
10^{12} M_\odot } {E_{\rm gas} \over E_{\rm b, gas}}\
\mu\hbox{K} \nonumber \, .
\label{DeltaT}
\end{eqnarray}

This SZ effect shows up on small (typically
sub-arcmin) angular scales. 

Quasar-driven blast-waves could inject into the ISM an amount of
energy several times higher than the gas binding energy, thus
producing larger, if much rarer, SZ signals.
A black-hole (BH) accreting a mass $M_{\rm BH}$ with a mass to
radiation conversion efficiency $\epsilon_{\rm BH}$ releases an
energy $E_{\rm BH}=\epsilon_{\rm BH}M_{\rm BH}c^2$. The
standard value for the efficiency $\epsilon_{\rm BH}=0.1$ and
a fraction $f_h=0.1$ for the energy fed in kinetic
form and generating strong shocks turning it into heat (see slide 85)
are assumed here for numerical estimates. 

Using the recent re-assessment by \cite{Tremaineetal2002} of
the well known correlation between the BH mass and the stellar
velocity dispersion,
%
%\begin{equation}
$M_{\rm BH} = 1.4 \times 10^8\,\left( {\sigma / 200\,{\rm 
km/s}}\right)^{4}\ 
{\rm M}_\odot$~,
%\ . \label{Mbh}
%\end{equation}
%
one gets
\begin{eqnarray}
{E_{\rm BH}\over E_{\rm b, gas}} 
%& 
\simeq 
%& 
4.7  \left({h\over 0.5}\right)^{-2/3}{\epsilon_{\rm BH}\over 0.1}\, {f_h\over 0.1} \,
{1+z_{\rm vir}\over 3.5}\,  
%\\
%&& 
\left({M_{\rm gas}/M_{\rm vir}\over0.1}\right)^{-1} \left({M_{\rm vir}\over 10^{12}
M_\odot}\right)^{-1/3} 
%\nonumber 
\, .  
\label{ratioT}
\end{eqnarray}
The amplitude of the SZ dip in the RJ region due to
quasar heating of the gas is then estimated as (see slide 86):
\begin{eqnarray}
%&& 
\left|\left({\Delta T \over T}\right)_{\rm RJ}\right|  \simeq 
1.8\times 10^{-5} {f_h \over 0.1}  
%\\
%&& 
\left({h\over 0.5}\right)^2
\left({\epsilon_{\rm BH} \over 0.1}\right)^{1/2} \left({E_{\rm
BH}\over 10^{62}}\right)^{1/2}\left({1+z \over 3.5}\right)^{3} 
%\nonumber 
\, . 
\label{DeltaTqso}
\end{eqnarray}
Following \cite{Plataniaetal2002}, an isothermal
density profile of the galaxy is adopted. The virial radius, encompassing a
mean density of $200\rho_u$, is then:
\begin{eqnarray}
R_{\rm vir} 
%& 
\simeq 
%& 
120 \left({h\over 0.5}\right)^{-1}
\left({E_{\rm BH}\over 10^{62}}\right)^{1/4}  
%\\
%&& 
\left({\epsilon_{\rm
BH}\over 0.1}\right)^{-1/4}\left(1+z_{\rm vir} \over
3.5\right)^{-3/2}\ \hbox{kpc} 
%\nonumber 
\, , 
\label{Rg}
\end{eqnarray}
corresponding to an angular radius:
\begin{eqnarray}
\theta_{SZ} 
%& 
\simeq 
%& 
17'' \left({E_{\rm BH}\over
10^{62}}\right)^{1/4} \left({\epsilon_{\rm BH}\over
0.1}\right)^{-1/4}
%  \\
%&& 
\left(1+z_{\rm vir} \over
3.5\right)^{-3/2}{d_A(2.5)\over d_A(z)} 
%\nonumber 
\, , 
\label{theta}
\end{eqnarray}
where $d_A(z)$ is the angular diameter distance.
                                                                                
The angular scales of these SZ signals from galaxies are of the order
of $\approx$ 10$^{\prime\prime}$, then of 
particular interest for a detailed mapping with the SKA and XEUS 
in the radio and X-ray,
respectively (see slide 87).
The probability of observing these SZ sources 
on a given sky field 
at a certain flux detection level  and the
corresponding fluctuations are mainly determined by 
the redshift dependent source number density 
$\phi_{\rm SZ}(S_{\rm SZ},z)$
per unit interval of the SZ 
(decrement) flux $S_{\rm SZ}$.
The lifetime of the considered SZ sources 
is crucial to determine their number density (see slide 88). 

For quasar-driven blast-waves the lifetime of the active phase,
$t_{\rm SZ}$, is approximately equal to the time for the shock
to reach the outer boundary of the host galaxy.
Assuming a self-similar blast-wave expanding in a medium with an
isothermal density profile, $\rho \propto r^{-2}$, one gets:
\begin{eqnarray}
t_{\rm SZ} 
%& 
\simeq 
%& 
1.5\times 10^8 \left({h\over 0.5}\right)^{-3/2}
\left({E_{\rm BH} \over
10^{62}\hbox{erg}}\right)^{1/8} 
% \\
%&& 
\left({\epsilon_{\rm BH}\over
0.1}\right)^{-5/8}\left({f_h\over
0.1}\right)^{-1/2}\left({1+z\over 3.5}\right)^{-9/4}\ \hbox{yr} 
%\nonumber 
\, .
\label{tSZ}
\end{eqnarray}

The evolving B-band luminosity function of quasars, $\phi(L_B,z)$, 
(see e.g. \cite{Pei1995})
can be then adopted to derive the 
source number density $\phi_{\rm SZ}(S_{\rm SZ},z)$ according to
\begin{equation}
\phi_{\rm SZ}(S_{\rm SZ},z) = \phi(L_B,z) {t_{\rm SZ} \over t_{\rm
q,opt}}\,{d L_B \over d S_{\rm SZ}} \ , \label{phiSZ}
\end{equation}
where $L_B(S_{\rm SZ},z)$ is the blue luminosity of a quasar at
redshift $z$ causing a (negative) SZ flux $S_{\rm SZ}$, 
and $t_{\rm q,opt}$
is the duration of the optically bright phase of the quasar
evolution (see slide 89). 

For the proto-galactic gas $t_{\rm SZ}$ should be replaced by the gas 
cooling time, $t_{\rm cool}$.
Assuming that quasars can be used as
effective signposts for massive spheroidal galaxies in their early
evolutionary phases \cite{Granatoetal2001} 
and that they emit at the Eddington limit
and using 
the relation by \cite{Ferrarese2002} between 
the mass of the dark-matter halo, $M_{\rm vir}$, and the 
mass of the central black-hole,
${M_{\rm BH}/ 10^8 M_\odot} \sim 0.1 
(M_{\rm vir} / 10^{12} M_\odot)^{1.65}$,
the number density of sources with gas at virial temperature
can be straightforwardly related to the quasar luminosity function
$\phi(L_B,z)$ (see slide 90). 

In spite of the many uncertainties of these models, it is remarkable
that the CMB fluctuations
(dominated at small scales by the Poisson contribution) 
induced by the SZ effect of these 
source populations could contribute to the CBI \cite{Masonetal2003} 
anisotropy measure and, in particular, could contribute to explain 
part of the excess in the angular power spectrum 
found by BIMA \cite{Dawsonetal2002} at multipoles 
$\ell \approx (4-10) \times 10^3$ (see slide 91),
a certain fraction of it being likely produced by
high-redshift dusty galaxies, 
whose fluctuations may be strongly enhanced by the effect of clustering
\cite{toffolatti05}.

Also, the integrated Comptonization distortion produced by quasar-driven blast-waves 
has been computed by \cite{Plataniaetal2002} through a quite general ``energetic''
approach, not particularly related to a detailed physical assumption: they found
$u \sim 2.4 \times 10^{-6}$.

%\begin{figure}[htb]
%\includegraphics[angle=0.,width=7.cm,height=7.cm]{vedi_clfromcf_2cases_add_da1000.ps}
%\caption{Angular power spectrum of SZ effects at 30~GHz
%compared to CMB primary fluctuation power spectrum and 
%CBI \cite{Masonetal2003} (box)
%and BIMA \cite{Dawsonetal2002} (data points)
%measures. Solid lines represent 
%clustering (bottom line), Poisson (middle line) and global 
%(upper line) contributions from quasar driven blast-waves.
%Dashed lines represent clustering (bottom
% line at high $\ell$), Poisson (middle line at high $\ell$) 
%and global (upper line) contributions from 
%proto-galactic gas. The latter are actually upper limits since,
% because of the uncertainty in the cooling time,
%the extreme assumption that $t_{\rm cool}=t_{\rm exp}$ has been adopted in 
%the computation.
%Dots refer to the overall contribution.} 
%\label{fig:sz_gal}
%\end{figure}

A direct probe of these models and, possibly, their accurate knowledge 
through a precise high resolution imaging is then 
of particular interest.
Slide 92 shows the number counts 
at 20~GHz predicted by these models:
in a single SKA FOV about few~$\times 10^2 - 10^3$ SZ sources with fluxes
above $\sim 100$~nJy could be then observed in few hours of 
integration.
Given the typical source sizes, 
we expect 
a blend of sources in the SKA FOV 
at these sensitivity levels, while much shorter integration times,
$\sim$~sec, on many FOV would allow to obtain much larger maps 
with a significant smaller number of resolved SZ sources per FOV.
Both surveys on relatively wide sky areas and deep exposures
on limited numbers of FOV are interesting and easily 
obtainable with SKA (see slide 93).

%\begin{figure}[htb]
%\includegraphics[angle=0.,width=7.cm,height=7.cm]{vedi_countsnu_2cases_forska.ps}
%\caption{Number count predictions at 20~GHz for SZ effects 
%as function of the absolute value of the flux 
%from proto-galactic gas heated at the virial temperature (dashes)
%assuming $M_{\rm gas}/M_{\rm vir} = 0.1$
%and from quasar driven blast-waves (solid line).
%The exponential model for the evolving luminosity function 
%of quasars is derived by
%\cite{Pei1995} 
%%for an Einstein-de Sitter universe, an Hubble
%%constant of $50\,\hbox{km}\,\hbox{s}^{-1}\,\hbox{Mpc}^{-1}$, 
%%and 
%for an optical spectral index of quasars $\alpha=0.5$ ($S_\nu \propto
%\nu^\alpha$). The parameters have been set at $\epsilon_{\rm
%BH}=0.1$, $f_h=0.1$, $k_{\rm bol}=10$, $t_{\rm q,opt}=10^7\,\hbox{yr}$.}
%\label{fig:n_sz_gal}
%\end{figure}

A different scenario to jointly explain the power excess 
found by BIMA and the high redshift reionization detected 
by WMAP 1-yr data \cite{kogutetal03} 
($z_{reion} \sim 15-20$) was proposed by \cite{OH03} (see slide 94). 
It involves hot gas winds powered by pair-instability supernovae (SN)
explosions from the first generation of very massive stars at very 
low metallicity able to photoevaporate the gas in the halo potential.
The SN remnants should then dissipate their energy in the 
intergalactic medium (IGM) and about 30-100\% of their energy  
would be transferred to the CMB via Compton cooling.
However, the resulting SZ effect from individual sources
is estimated to be too faint (corresponding to $\approx {\rm few} \times 
10^{-2}$~nJy) to be observable even by SKA.
On the contrary, the resulting SZ effect from these sources could be 
relevant in statistical sense. It is claimed to explain 
the high $\ell$ BIMA excess of the CMB angular power spectrum
and to be able to generate a global Comptonization distortion
parameter $u \sim {\rm few} \times 10^{-6}$ (on the other hand,
a certain decrease of its effect with respect to this picture
could be expected in the light of the lowering of the reionization 
redshift as revised by WMAP 3-yr data).

\subsection{Free-free emitters}
\label{ff_emitt}

The understanding of the ionizing emissivity of collapsed objects and 
the degree of gas clumping is crucial for reionization models.
The observation of diffuse gas and Population III objects
in thermal bremsstrahlung as a direct
probe of these quantities has been investigated by 
\cite{OH99}. Free-free emission produces both global 
and localized spectral distortion of the CMB (see slide 95).
A natural way to distinguish between free-free distortion 
by ionized halos rather than by diffuse ionized IGM
is represented by observations at high resolution of dedicated sky areas 
and 
by the fluctuations in the free-free background.
In the model by \cite{OH99}  halos collapse and form
a starburst lasting $t_{o}=10^{7}$ yr, then recombine and no longer
contribute to the free-free background.
By adopting a Press-Schechter model 
\cite{PressSchechter1974,Bondetal1991}
for the number density of collapsed 
halos per mass interval, $d n_{PS}/dM$, \cite{OH99} 
exploited the expression by \cite{sasaki1994}
for the collapse rate of halos per mass interval per unit comoving volume
(see slide 96):
\begin{equation}
\frac{d\dot{N}^{form}}{dM}(M,z)=
\frac{1}{D}\frac{dD}{dt}\frac{d n_{PS}}{dM}(M,z)
\frac{\delta_{c}^{2}}{\sigma^{2}(M)D^{2}} \, ; 
\end{equation}
here
$D(z)$ is the growth factor and $\delta_{c}=1.7$ is the
threshold above which mass fluctuations collapse.
The expected comoving number density of ionized halos in a given flux
interval as a function of redshift 
\begin{equation}
\frac{dN_{\rm halo}}{dS dV}(S,z)= \int_{t(z)}^{t(z)-t_{o}} dt
\frac{d \dot{N}^{form}}{dM} \frac{dM}{dS}
\end{equation}
can be then computed given the expected flux from a halo of mass M at 
redshift $z$,  $S=S(M,z)$, and the starburst duration, $t_{o}$.
Adopting a cut-off mass for a halo (see slide 97)
to be ionized of
$M_{*}=10^{8} (1+z/10)^{-3/2} M_{\odot}$ 
(the critical mass needed to attain a virial temperature 
of $10^{4}$~K to excite atomic hydrogen cooling), 
\cite{OH99} computed the number counts of sources above the flux limit 
$S_{c}$ from the zeroth
moment of the intensity distribution 
moments due to sources
above a redshift $z_{\rm min}$,
\begin{eqnarray}
%&& 
\langle S^{n}(>z_{min},S_{c}) \rangle 
%\\
%&& 
 = \int_{z_{\rm min}}^{\infty} dz
\int_{S_{\rm min}(z)}^{S_{max}} dS
\frac{dN_{\rm halo}}{dSdV} \frac{dV}{dz d\Omega} S^{n} 
%\nonumber 
\, ,
\label{moments}
\end{eqnarray}
by setting $S_{max} \rightarrow \infty$ 
and $S_{min}(z)={\rm max}(S_{c},S_{*}(z))$, where 
$S_{*}(z)$ denotes the flux from a halo of
minimum mass $M_{*}$ at redshift $z$.

The relation 
%\begin{equation}
$\dot{N}_{\rm recomb}= \alpha_{B} \langle n_{e}^{2} \rangle V \approx
(1-f_{esc})\dot{N}_{\rm ion} \, ,$
%\label{nesq}
%\end{equation}
between the production rate of
recombination line photons, $\dot{N}_{\rm recomb}$, 
and the production rate of ionizing photons,
$\dot{N}_{\rm ion}$,
(here $\alpha_{B}$ is the 
recombination coefficient 
and $f_{esc}$ ($\approx {\rm some} \%$) is the escape fraction for 
ionizing photons)
implies that
the source luminosities in H$\alpha$ and free-free emission 
($\propto n_{e}^{2}V$)
are proportional to the production rate of ionizing photons (see slide 98).
Over a wide range of nebulosity conditions \cite{HummerStorey1987} 
found that $\simeq$~0.45~H$\alpha$ photons are emitted per Lyman 
continuum photon; thus 
${\rm L(H \alpha)} = 1.4 \times 10^{41}
{\dot{N}_{\rm ion}}/({10^{53} {\rm
ph \, s^{-1}}})  {\rm erg \, s^{-1}}$. 
Given the free-free volume
emissivity
\cite{Rybicki_Lightman_1979}
in the case of an approximate mild temperature 
dependence with a power law (a velocity averaged Gaunt factor
$\bar{g}_{ff}=4.7$ is assumed),
$\epsilon_{\nu} = 3.2 \times 10^{-39} n_{e}^{2} ({T}/{10^{4} K})^{-0.35}
\ {\rm erg \,s^{-1} \,cm^{-3} \, 
Hz^{-1}}$~,
it is found
\begin{equation}
L_{\nu}^{ff}=1.2 \times 10^{27} \frac{\dot{N}_{\rm
ion}}{10^{53} {\rm ph \, s^{-1}}} {\rm erg \, s^{-1} \,
Hz^{-1}} \, .
\label{L_free}
\end{equation}
Assuming the starburst model of \cite{HaimanLoeb1998} 
normalised to the observed metallicity
$ 10^{-3} Z_{\odot} \le Z \le  10^{-2} Z_{\odot}$ of
the IGM at $z \sim 3$ (resulting into 
a constant fraction of the gas mass turning into stars,
$1.7 \% \le f_{star} \le 17\%$,
in a starburst which fades after $\sim 10^{7}$~yr),  \cite{OH99} 
derived  a production rate of ionizing photons as a function of halo mass
given by (see slide 99):
\begin{equation}
\dot{N}_{\rm ion}(M)=2 \times 10^{53} \,
{f_{star} \over 0.17} \, {M \over 10^{9} M_{\odot}} 
\, {\rm ph} \,  {\rm s^{-1}} \, ,
\label{Ndot_scaling}
\end{equation}
which specifies the above free-free ionized halo
luminosity. The corresponding flux is then:
\begin{eqnarray}
%&& 
S_{\rm ff} = \frac{L_{\nu}^{\rm ff}}{4 \pi d_{L}^{2}}
(1+z)
% \\
%& 
\approx
% & 
2.5 \left( \frac{1+z}{10} \right)^{-1}
\frac{\rm M} {10^{9} {\rm M_{\odot}}}
\left( \frac{\rm T} {10^{4} \, {\rm K}} \right)^{-0.35} {\rm nJy} 
%\nonumber 
\, .
\label{J_free_free}
\end{eqnarray}

%\begin{figure}[htb]
%\includegraphics[angle=0.,width=7.cm,height=7.cm]{c_Num_sources_free.ps}
%\caption{Number of sources which may be detected in the $1^{\circ}$
%by SKA, as a function of the threshold flux
%$S_{c}$. Realistic limiting fluxes for point source detection are shown. 
%The extrapolated source counts from \cite{Partridgeetal1997} are also 
%shown.
%From \cite{OH99}.}
%\label{Num_sources_free}
%\end{figure}

Clearly, SKA will allow to detect only bright sources
with deep exposures. 
The ionized halo number counts can be calculated from
Eq.~(\ref{moments}).
The result by \cite{OH99} is reported in 
slide 100 (and compared with \cite{Partridgeetal1997}):
SKA should be able to detect $\sim 10^{4}$
individual free-free emission sources with $z>5$ 
in 1 square degree above a source detection threshold of 70~nJy. 
The redshift information from the Balmer line emission detectable
by the Next Generation Space Telescope (NGST) can be used to discriminate
ionized halos from other classes of radio sources.

Ionized halos may contribute 
to the temperature fluctuations. In particular, the Poisson contribution
is predicted to be larger (smaller) than the clustering one at  
scales smaller (larger) than $\sim 30''$ \cite{OH99}. On the other hand,
both are likely dominated by the radio source contribution. 

Finally, the integrated emission from ionized halos produces a global CMB 
spectral distortion,
$\Delta T_{ff}=c^{2}\langle S \rangle /2k_B\nu^{2}$, 
that can be computed from the 
mean sky averaged signal $\langle S \rangle$ (see slide 101).
By using Eq.~(\ref{moments})
(since no point source removal is feasible at degree scales) 
with $z_{\rm min}$ and $S_{c}=0$, 
\cite{OH99} found a free-free distortion 
$\Delta T_{ff}= 3.4 \times 10^{-3}$~K at 2~GHz, corresponding to 
a free-free distortion parameter $y_{B}\simeq 1.5\times 10^{-6}$,
well within the observational capability of a next generation of
CMB spectrum experiments at long wavelengths (see next section).

\subsection{SKA contribution to future CMB spectrum experiments}
\label{cmbspec}

The current limits on CMB spectral distortions and the constraints on 
energy dissipation processes $|\Delta \epsilon / \epsilon_i| \lsim 
10^{-4}$ in the plasma \cite{SB02}
are mainly set by COBE/FIRAS \cite{mather90,fixsen96} (see slides 102--104).
CMB spectrum experiments from space,
DIMES \cite{KOG96} (see also \cite{KOG03})
at $\lambda \gsim 1$~cm and FIRAS~II \cite{FM02}
at $\lambda \lsim 1$~cm, 
have been proposed with an accuracy potentially able to 
constrain (or probably detect) energy exchanges 10--100 times 
smaller than the FIRAS upper limits (see slide 105).
Also, a promising opportunity in this respect is represented by the recent 
renaissance of interest for the Moon as a base for extremely 
accurate observations of the universe. Slides 111--113 display some
of the fundamental ideas for a CMB spectrum experiment at long wavelengths,
proposed in a recent study for ASI, that forsee an ultimate huge
experiment at centimetre and decimetre wavelengths anticipated by 
precursor experiments at centimetre wavelengths, able to significantly improve
the current observational scenario \cite{luna}. 

Long wavelength experiments may probe in particular dissipation
processes at early times ($z \gsim 10^5$)
resulting in Bose-Einstein like distortions 
\cite{SZ70,DD80,BDD91a} 
and free-free 
distortions \cite{bart_stebb_1991}
possibly generated by heating (but, although disfavoured
by WMAP, in principle  
also by cooling \cite{stebb_silk}) 
mechanisms at late epochs ($z \lsim 10^4$),
before or after the recombination era \cite{BS03a}.

Typical shapes of distorted spectra 
potentially detectable with new experiments
are shown in slide 108 while slides 109 and 110 display the possible improvements in our
understanding of the energy exchange in the primeval plasma 
respectively in the absence and in the presence
of detection of spectral distortions. 
To firmly observe such small
distortions the Galactic and extragalactic foreground contribution
should be accurately modelled and subtracted (see slide 114). 
%\begin{figure}[htb]
%\includegraphics[angle=0.,width=7.cm,height=7.cm]{all_dist.ps}
%\caption{CMB distorted spectra as functions of the wavelength
%$\lambda$ (in cm) in the presence of
%a late energy injection with $\Delta \epsilon / \epsilon_i = 5 \times 
%10^{-6}$ plus an early/intermediate
%energy injection with $\Delta \epsilon / \epsilon_i = 5 \times 10^{-6}$
%occurring at the ``time'' Comptonization 
%parameter
%$y_h=5, 1, 0.01$ (from the bottom to the top;
%in the figure the cases at $y_h=5$ and 1 are indistiguishable
%at short wavelengths; solid lines) and plus a
%free-free distortion
%with  $y_B=10^{-6}$ (dashes).
%$y_h$ is defined by Eq.~(4) 
%with $dl=cdt$ and $T_e=T_{CMB}$ when the integral
%is computed from the time of the energy injection
%to the current time.} 
%\label{fig:cmb_dist}
%\end{figure}
Recent progress on radio source counts at 1.4~GHz have been presented
in \cite{prandonietal01}. On the other hand, the very faint tail
of radio source counts is essentially unexplored and their 
contribution to the radio background at very low brightness 
temperature is not accurately known (see slide 114).   
For illustration, by assuming differential source number counts,
$N(S)$, given by ${\rm log} N(S)/\Delta N_0 \sim a {\rm log} S + b$,
with $\Delta N_0 \sim 150 S^{-2.5}$~sr$^{-1}$~Jy$^{-1}$ ($S$ in Jy)
\cite{toffolattietal98},
for $a \sim 0.4-0.6$ and $b \sim - (0.5-1)$, one derives a contribution
to the radio background at 5~GHz 
from sources between $\sim 1$~nJy and $\sim 1 \mu$Jy 
between few tens of $\mu$K and few mK (see slide 115).
These signals are clearly negligible compared to the accuracy of current 
CMB spectrum experiments, in particular at $\lambda \gsim 1$~cm,
but are significant at the accuracy level on CMB distortion 
parameters potentially achievable with novel experiments at long wavelengths.
This effect is small compared to the Galactic 
radio emission, whose accurate knowledge currently represents 
the major astrophysical problem in CMB spectrum experiments,
but, differently from Galactic emission, it is 
isotropic at the angular scales of few degrees 
and can not be then subtracted from the CMB monopole 
temperature on the basis of its angular correlation properties. 
With accurate absolute measures on a wide frequency coverage
a fit including both CMB distorted spectra and astrophysical 
contributions can be searched (see \cite{SB02} for an application
to FIRAS data) but  a direct radio background estimate
from precise number counts will certainly improve the robustness
of this kind of analyses.      

The SKA sensitivity at 20~GHz will allow 
the detection (to $5\sigma$) 
of sources down to a flux level of 
$\simeq 200$~nJy
($\simeq 60, 20, 6$~nJy) 
in 1 (10, $10^2$, $10^3$) hour(s) of integration 
over the $\simeq 1$~mas (FWHM) resolution element;
similar numbers (from $\simeq 250$ to 8~nJy 
in an integration time from 1 to $10^3$ hours,
respectively) but on a resolution element about 10 times
larger will be reached at $\approx$~GHz frequencies
by using a frequency bandwidth of about 25\%.

Therefore, the SKA accurate determination of source number 
counts down to very faint fluxes can directly help the solution of
one fundamental problem of the future generation of CMB 
spectrum space experiments at $\lambda \gsim 1$~cm (see slide 116).

\section{Conclusion}

The thermal plasma in the intergalactic and intracluster medium and at
galactic scales leaves imprints on the CMB 
through the Thomson scattering
of CMB photons on  hot electrons (SZ effect) 
and the free-free emission.
Since its original formulation, the SZ effect, first elaborated 
for clusters of galaxies,
has been recognized as a ``powerful laboratory'' for our comprehension 
of physical processes in cosmic structures and to derive crucial
information on some general properties of the universe. 

Many thousands of galaxy clusters
can be identified by XMM, {\it Planck}, and SDSS. 
SKA will allow to map the thermal and density 
structure of clusters of galaxies at radio and centimetre bands
with unprecedented resolution and sensitivity 
and with an extremely accurate control of 
extragalactic radio source contamination.
The signatures from SZ effects and free-free emission
at galactic scales and in the intergalactic medium probe
the structure evolution at various cosmic times. 
The detection of these sources and their
imaging at the high resolution and sensitivity achievable 
with SKA will greatly contribute 
to the comprehension of crucial cosmological and astrophysical aspects, 
as the physical conditions of early ionized halos, quasars
and proto-galactic gas.
A wealth of information will be available by combining SKA observations with
those expected at higher frequencies by ALMA and NGST and with the 
high quality X-ray data promised by XEUS. 
Also, the spectacular improvement in our understanding of the properties 
of extragalactic radio sources at very faint fluxes 
achievable with SKA will allow to accurately model their contribution 
to the diffuse background from radio to centimetre wavelengths and
will greatly contribute to the interpretation of 
next generation of CMB spectrum experiments.

In conclusion, although not specifically devoted to CMB studies, because of its 
high resolution and the limited high frequency coverage, 
the extreme sensitivity and resolution of SKA may be fruitfully 
used for a detailed mapping of the 
thermal plasma properties in the 
intergalactic and intracluster medium and at
galaxy scale and to probe the
thermal plasma history at early times.

\vskip 1cm

\noindent
{\bf Acknowledgements --} I warmly thank G. De Zotti and L. Feretti for 
numberless and constructive conversations and for the collaboration 
to the SKA scientific case.
It it a pleasure to thank the School organizers for the kind invitation 
and generous and helpful logistic support.
The use of the {\sc cmbfast} code\footnote{http://www.cmbfast.org/}
(see e.g. \cite{seliak_zalda_96}) is acknowledged.

\end{document}